\documentclass[aps,twocolumn,prb,superscriptaddress]{revtex4-2}
\usepackage[utf8]{inputenc}

\usepackage{natbib}
\usepackage{graphicx}
\usepackage{dcolumn}
\usepackage{bm}
\usepackage{epsfig}
\usepackage{textcomp}
\usepackage{gensymb}
\usepackage{xcolor}
\usepackage{placeins}
\usepackage{physics}
\usepackage{amssymb,amsmath,amsthm,enumitem}
\usepackage{float}

\usepackage{comment}
\usepackage{caption}
\usepackage{subcaption}
\usepackage{booktabs}
\usepackage{lipsum}
\usepackage{placeins}
\usepackage{ragged2e}
\usepackage{makecell}

\begin{document}
\title{Exchange field induced symmetry breaking in quantum hexaborides}
\author{D. Rivera}
\affiliation{Instituto de Física, Universidade de São Paulo,  05315-970 São Paulo, São Paulo, Brasil}

\author{Fernando P. Sabino}
\affiliation{Departamento de Engenharia de Materiais, Universidade de São Paulo, 13563-120 São Carlos, São Paulo, Brasil}

\author{H. Raebiger}
\affiliation{Department of Physics, Yokohama National University, Yokohama, Japan}

\author{A. Ruzsinszky}
\affiliation{Department of Physics and Engineering Physics, Tulane University, New Orleans, LA 70118}

\author{J. P. Perdew}
\affiliation{Department of Physics and Engineering Physics, Tulane University, New Orleans, LA 70118}

\author{G. M. Dalpian}
\email{dalpian@usp.br}
\affiliation{Instituto de Física, Universidade de São Paulo,  05315-970 São Paulo, São Paulo, Brasil}

\begin{abstract}
Symmetry breaking (SB) has proven to be a powerful approach for describing quantum materials: strong correlation, mass renormalization, and complex phase transitions are among the phenomena that SB can capture, even when coupled to a mean-field-like theory. Traditionally, corrective schemes were required to account for these effects; however, SB has emerged as an alternative that can also successfully describe the intricate physics of quantum materials. Here, we explore spin SB on EuB$_6$ and SmB$_6$ and how its relation to the exchange field can determine onsite properties, depending on the type of symmetry breaking. Using spin-polarized Density Functional Theory (DFT) calculations with the $r^2$SCAN functional, we systematically compare four magnetic configurations, one totally symmetric - non-magnetic (NM) configuration - and three with different types of symmetry breaking: ferromagnetic (FM), antiferromagnetic (AFM) and a paramagnetic (PM) configuration - modeled through a Special Quasirandom Structure (SQS) method - to capture local symmetry-breaking effects. Our results show that the PM configuration produces distinct magnetic environments for the rare-earth atoms, leading to different exchange fields. These, in turn, induce symmetry breaking in the electronic and magnetic properties of Eu and Sm. Those results provide an alternative explanation for the experimental results on both materials, EuB$_6$ and SmB$_6$, where X-ray Absorption Spectroscopy (XAS) and X-ray Absorption Near Edge Structure (XANES) measurements suggest the presence of multiple atomic environments, previously attributed to a mixed-valence configuration.
\end{abstract}

\maketitle

\section{Introduction}
\label{sec:intro}

Symmetry breaking (SB) is a recurring and foundational concept in materials physics \cite{Anderson1972}. A classic example is the Jahn-Teller distortion, where a lifting of electronic degeneracy drives a structural distortion, altering the symmetry of the lattice and impacting optical, magnetic, and dynamical properties \cite{jahntellet1937,dresselhaus2007group}. Another widely studied case involves charge density waves (CDWs), in which the translational symmetry of the crystal is broken by the formation of a modulated charge pattern \cite{Miao2021}. While such mechanisms have been explored for decades, novel instances of symmetry breaking continue to emerge, often revealing unexpected physical phenomena \cite{Zunger2022,zunger2020,zunger2025-3,Joshi2024,Liu2025,perdew2022}. In particular, breaking the symmetry of a \textit{motif} can uncover rich material-specific physics, opening new directions in the study of quantum matter \cite{Zunger2022}. 

A motif is a pattern of a microscopic Degree of Freedom (mDOF) within a system, such as structural, spin, or dipole motifs \cite{Zunger2022, Jiang2022, Malyi2020}. Ferromagnetic (FM) and ferroelectric (FE) configurations are examples of spin and dipole motifs, respectively.
Within this framework, three symmetry-based configurations can be defined: monomorphous (local and global symmetries preserved), polymorphous (local symmetry broken, global preserved), and long-range ordered (LRO) states \cite{Zunger2022}. This motif-based SB approach has been successfully applied to various quantum materials \cite{Malyi2021, Malyi2022, Malyi2023,  Joshi2024, Xiong2025}. For spin systems, NM, PM, and FM/AFM states exemplify these configurations, respectively. Because the spin degree of freedom is crucial for the electronic behavior in many correlated materials \cite{zunger2020,Sun2018,Malyi2021,chen2024,Jiang2024}, applying SB to the spin channel yields not only quantitative renormalizations but can also drive qualitative modifications in material behavior and stabilize long-range magnetic order.

To evaluate spin-SB in a deeper manner, we begin with the spin-resolved Kohn–Sham (KS) Hamiltonian within density functional theory (DFT), which provides the formal framework for introducing and understanding symmetry breaking in the spin channel \cite{giustino2014,engel2011density}:
\begin{equation}
    \label{eq:spin-KS_equation}
    \left[ -\frac{\nabla^2}{2} + v_{eff}(\boldsymbol{r}) + \mu_B \boldsymbol{\sigma} \cdot \boldsymbol{B}_{xc} (\boldsymbol{r}) \right] \boldsymbol{\Psi}_i(\boldsymbol{r}) = \epsilon_{i\sigma} \boldsymbol{\Psi}_i(\boldsymbol{r}),
\end{equation}
with
\begin{equation*}
    v_{eff}(\boldsymbol{r}) = v_{ext}(\boldsymbol{r}) + v_H(\boldsymbol{r}) + v_{xc}(\boldsymbol{r}).
\end{equation*}

\noindent Here $\boldsymbol{\Psi}_i$ is the 2 component KS spinor ($\uparrow$ and $\downarrow$ states) \cite{giustino2014}, which is used to construct the vector spin densities $\boldsymbol{s}(\boldsymbol{r})$, and $\epsilon_{i\sigma}$ are the respective spin-resolved KS eigenvalues. Besides, $v_{eff}$ represents an effective potential describing the electron interaction with ions, electron density (classically) and other electrons (quantum) respectively \cite{martin2004electronic,giustino2014}. Finally, the last term is the coupling between the electron spin with the exchange-correlation magnetic field $\boldsymbol{B}_{xc} (\boldsymbol{r})$, or exchange field for short \cite{Krawczyk2012}.

Formally, $\boldsymbol{B}_{xc}$ is defined as the functional derivative of the exchange-correlation energy with respect to the spin density \cite{Sharma2018}, giving it the character of a vector field \cite{Heisenberg1928}. It can be viewed as an internal magnetic field arising from atomic magnetic moments \cite{giustino2014}, or as a mean-field representation of the exchange interaction \cite{Krawczyk2012,Heisenberg1928}, but including correlation effects. 

The concept of an exchange field predates the development of Density Functional Theory. In his seminal 1928 paper, Heisenberg showed that the Weiss \textit{molecular field}—an internal field responsible for spin alignment in ferromagnets—originates from quantum-mechanical exchange interactions, \textit{i.e.} a combination of coulomb interactions and the Pauli exclusion principle for fermions \cite{Heisenberg1928}. Later, Slater extensively discussed the concept of an exchange field in several works, providing key insights that helped shape modern theoretical approaches \cite{Slater1936,Slater1951}. In this way, the consideration of the exchange field in magnetic environments is, and has always been, crucial.

Since both $\boldsymbol{B}_{xc}(\boldsymbol{r})$ and $v_{xc}(\boldsymbol{r})$ depend on the electron and spin densities through $E_{xc}[n(\boldsymbol{r}),\boldsymbol{s}(\boldsymbol{r})]$, breaking the spin motif symmetry directly alters the Hamiltonian. Recent studies emphasize the impact of such symmetry breaking in strongly correlated systems. Notably, Ullrich's study on a four-site Hubbard model \cite{Pluhar2019} shows that breaking $\boldsymbol{B}_{xc}$ symmetry lowers the total energy, making the symmetry-broken state energetically favorable. In the strong correlation limit, this leads to substantial site-dependent variations in magnetization and charge density between symmetric and symmetry-broken solutions. These findings highlight the fundamental role of local spin symmetry breaking in defining electronic and magnetic properties, reinforcing its importance in realistic calculations.

Magnetic clusters and spin polarons \cite{Bondarenko2017,spin_polaron_eub6-2,vanKooten2024} generate different magnetic environments, effectively lowering the symmetry of $\boldsymbol{B}_{xc}$, $v_{xc}$, and the Kohn–Sham Hamiltonian $H_{KS}$ as well, enhancing its local character and impacting atomic-scale properties. This theoretical framework is supported by studies such as those by Malyi \textit{et al.}, who observed changes in the local magnetic moment distribution in $EuTiO_3$ under the polymorphic configuration (PM) \cite{Malyi2022}, and by Zhi Wang \textit{et al.}, who reported similar effects in $FeSe$ \cite{Xhi_Whang2020}, consistent with a symmetry broken exchange field affecting local electronic and magnetic behavior.

Signatures of coexisting atomic environments—possibly reflecting magnetic polymorphic phases—have been experimentally reported for several materials, including the quantum hexaborides EuB$_6$ and SmB$_6$. In EuB$_6$, X-ray Absorption Near Edge Structure (XANES) measurements under high hydrostatic pressure reveal the emergence of two distinct atomic environments for Eu atoms. These environments were interpreted as a manifestation of mixed valence, involving both Eu$^{2+}$ and Eu$^{3+}$ states, despite the absence of a structural phase transition \cite{Donizeth2023}. This is accompanied by a collapse of magnetic ordering and suggests the stabilization of a new magnetic phase, either AFM or PM \cite{Donizeth2023}. Similarly, SmB$_6$, known for its non-magnetic ground state with mixed-valence Sm atoms observed via X-ray Absorption Spectroscopy (XAS), displays localized magnetic moments with short-range correlations \cite{PhysRevB.105.195134}. Under high pressure, the mixed valence state vanishes \cite{Chen2018}, and the system undergoes a transition into a magnetically ordered phase \cite{Barla2005}.

Understanding why symmetry breaking tends to well describe some quantum materials is still a mystery. One interpretation is that symmetry breaking transforms strong correlation in a symmetric state into normal correlation that density functional approximations can describe \cite{perdew2022,Perdew2021}, and is a path toward the classical limit \cite{Perdew2025}. As Zunger emphasized, combining symmetry-breaking (SB) calculations with local experimental techniques is essential to advancing our understanding of quantum materials \cite{Zunger2022}. This approach enables us to assess whether SB plays a fundamental role in their behavior.

In this work, we apply the concept of spin SB to investigate the fundamental differences between the monomorphous NM state, LRO magnetic phases—such as FM and AFM—and polymorphous (PM) configurations. SB calculations require an increase in the system’s degrees of freedom, typically achieved through the use of supercells \cite{Zunger2022}. We focus on the quantum hexaborides EuB$_6$ and SmB$_6$, where local probes like XANES and XAS \cite{zunger2025-3} have revealed different atomic environments, making them ideal for exploring the role of SB in correlated systems. In PM phases, spin SB leads to site-dependent exchange fields from distinct local spin arrangements, creating two unequivalent rare-earth atomic environments. This could be an alternative explanation for the experimental results in both materials. Notably, our calculations show no evidence of mixed valence in either material, indicating that these SB-induced effects arise without changes in valence. SB manifestations are found to be more pronounced in SmB$_6$ than in EuB$_6$.

\begin{figure*}[ht]
    \centering
    \includegraphics[scale = 0.28]{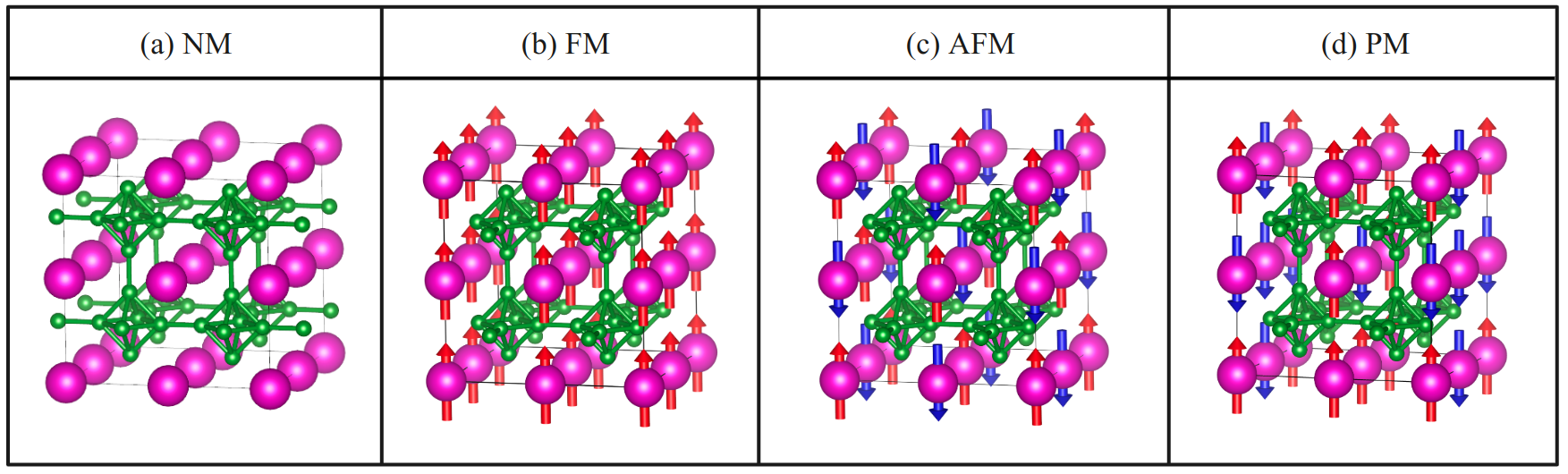}\
    \caption{\justifying Crystal structure of the hexaborides (SmB$_6$ or EuB$_6$) for the (a) NM, (b) FM, (c) AFM, and (d) PM configurations. Red are spin-up states, and blue arrows are spin-down states.}
    \label{fig:configurations}
\end{figure*}

\section{Computational methods}

To study symmetry breaking in quantum hexaborides we have used quantum mechanical methods based on the spin Density Functional Theory (spin-DFT) \cite{kohn1964,Kohn1965}. The $r^2SCAN$ meta-GGA functional was used to recover exchange and correlation effects \cite{Furness2020}. Calculations were performed using the Projector-Augmented Wave (PAW) method \cite{paw}, implemented in the Vienna Ab-initio Simulation Package (VASP) software \cite{vasp1,vasp2}. Projectors were chosen to include all $f$ electrons for every calculation within the PAW formalism. The energy cutoff was set to $400\: eV$. A $2 \times 2 \times 2$ supercell was created, as shown in figure \ref{fig:configurations}. We have calculated the most stable structural configurations by relaxing both lattice parameters and atomic positions for every system in each magnetic configuration. Electronic convergence was set to occur on a total energy tolerance of $1 \times 10^{-5}\: eV$, and forces were considered to converge if the magnitudes on individual atoms were below $0.01 \:\: eV/$\AA. For both EuB$_6$ and SmB$_6$ a gamma-centered $5 \times 5 \times 5$ k-mesh was used for ionic relaxation and to calculate the Densities Of States (DOS's). Bands are unfolded due to the nature of the $2 \times 2 \times 2$ supercell, and \textit{vaspkit} was used to perform the band unfolding procedure \cite{vaspkit}. NM, FM, AFM and PM magnetic configurations were studied, as shown in figures \ref{fig:configurations} (a), \ref{fig:configurations} (b), \ref{fig:configurations} (c) and \ref{fig:configurations} (d). The PM configuration was created using the SQS method, implemented in the ATAT software, using a Monte Carlo approach \cite{Zunger1990,VANDEWALLE2002539}. Due to the supercell size, using doublets or triplets correlation functions resulted in the same SQS structure, shown in figure \ref{fig:configurations} (d). The SQS method was used to introduce spin disorder only on Eu/Sm atoms, since they are the magnetic species, creating the polymorphous model.

In our work, the spin-symmetry breaking occurs spontaneously from the non-magnetic state. We tried "nudging" lattice-symmetry breaking and mixed-valence symmetry breaking, and did not find it. However, not finding something does not imply that it is not there.

\section{Results}

\subsection{Monomorphous Configuration}

The representative state of the monomorphous configuration is the non-magnetic (NM) state. It means the electron's wavefunctions are scalar objects and every state is doubly occupied, reflecting that spin-up and spin-down respect SU(2) symmetry, \textit{i.e.} any rotation in the spin space does not alter the physics of the materials. We have calculated the band structure for each material in the NM configuration, as shown in figure \ref{fig:bands_nm}. In this case, all Eu/Sm atoms are equivalent, \textit{i.e.} they experience the same magnetic environment, which is none. Thus, no exchange field is felt by the atoms within the system. We can see flat $f$-bands just below and just above the Fermi level, suggesting the possibility of strong correlation in the symmetric or monomorphous state.

\begin{center}
\includegraphics[scale = 0.48]{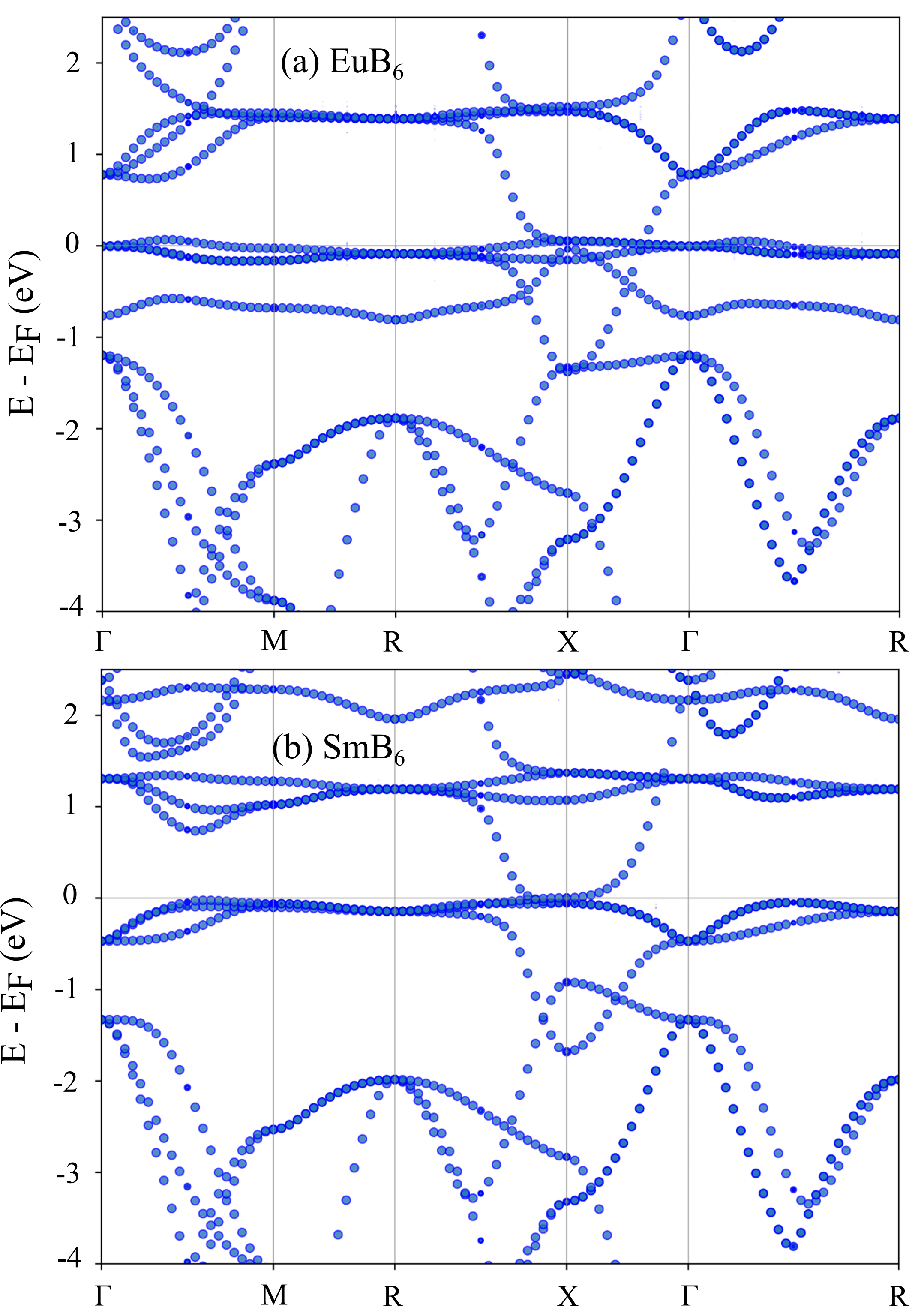}
\captionof{figure}{Unfolded band structures for the NM monomorphous configuration. Panel (a) shows the results for EuB$_6$, while (b) illustrates SmB$_6$. The spin-up and spin-down bands are identical.}
\label{fig:bands_nm}
\end{center}

\subsection{Long-range order (LRO) configurations}

The FM phase is the ground state magnetic configuration for EuB$_6$, as discussed in the introduction. For SmB$_6$, our calculations indicate that the PM configuration is the most stable among the magnetic phases we explored. In these magnetic states, the system exhibits a spontaneous breaking of spin-rotation symmetry: the degeneracy between spin-up and spin-down states is lifted, meaning that rotations in spin space no longer leave the physics invariant. The relative energy between these configurations are small, suggesting that multiple magnetic configurations—such as FM and AFM arrangements—are energetically competitive. This behavior is consistent with previous reports in the literature~\cite{Zhang2022}. A detailed summary of the relative energies and the average magnetic moments of the rare-earth ions is provided in Table~\ref{tab:magnetic_configurations}.
\begin{table}[h]
    \centering
    \caption{\justifying Relative energies and local magnetic moments for the different magnetic configurations of EuB$_6$ and SmB$_6$.}
    \label{tab:magnetic_configurations}
    \begin{tabular}{llcc}
        \hline
        Material & Config. & \makecell{ Relative \\ energy (meV/atom)} & Local MM ($\mu_{\text{B}}$) \\
        \hline
        EuB$_6$ 
            & FM  &  0 & 6.670 \\
            & PM  &  +0.47 & 6.662 \\
            & AFM &  +0.79 & 6.661 \\
            & NM  &  +1138.14 & 0.000 \\
        \hline
        SmB$_6$
            & PM  &  0 & 5.494  \\
            & FM  &  +3.01 & 5.521 \\
            & AFM &  +5.75 & 5.500 \\
            & NM  &  +830.98 & 0.000 \\
        \hline
    \end{tabular}
\end{table}

Spin symmetry breaking (SB) plays a stabilizing role in both EuB$_6$ and SmB$_6$. The monomorphous nonmagnetic (NM) configuration lies significantly higher in energy compared to any configuration that allows for spin SB. This indicates that spin SB is essential for accurately describing the physics of both quantum hexaborides—even in the paramagnetic (PM) phase—challenging the traditional paradigm of representing PM states through monomorphous, symmetry-constrained structures \cite{Zunger2022}. Instead, our results support the view that symmetry breaking at the local level, even in the absence of long-range magnetic order, captures key aspects of the electronic and magnetic behavior of these materials. Strong thermal mixing of the broken-symmetry states is to be expected, as in Ref. \cite{Wang2008}.

Figure~\ref{fig:bands_fm} displays the FM band structures of both EuB$_6$ and SmB$_6$. While FM is not the ground state of SmB$_6$, this configuration remains of physical interest because of its small relative energy.
\begin{center}
\includegraphics[scale = 0.51]{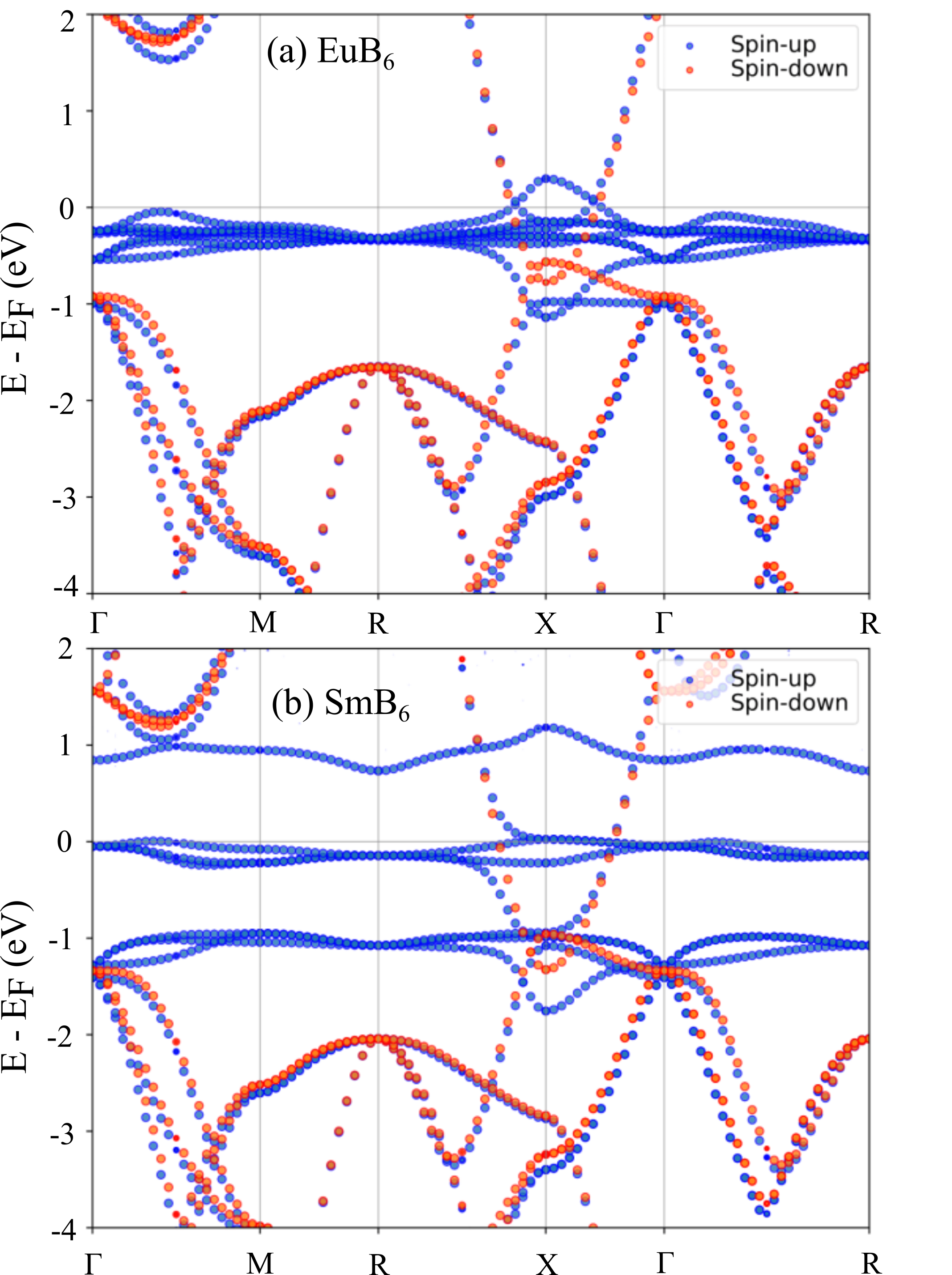}
\captionof{figure}{Unfolded band structures for the FM configuration. Panel (a) shows the results for $EuB_6$, while (b) illustrates $SmB_6$.}
\label{fig:bands_fm}
\end{center}

The AFM configuration exhibits similar features as the FM one. However, in this case the spin-up and spin-down states are fully degenerate, reflecting the absence of exchange splitting.
\begin{center}
\includegraphics[width=0.485\textwidth]{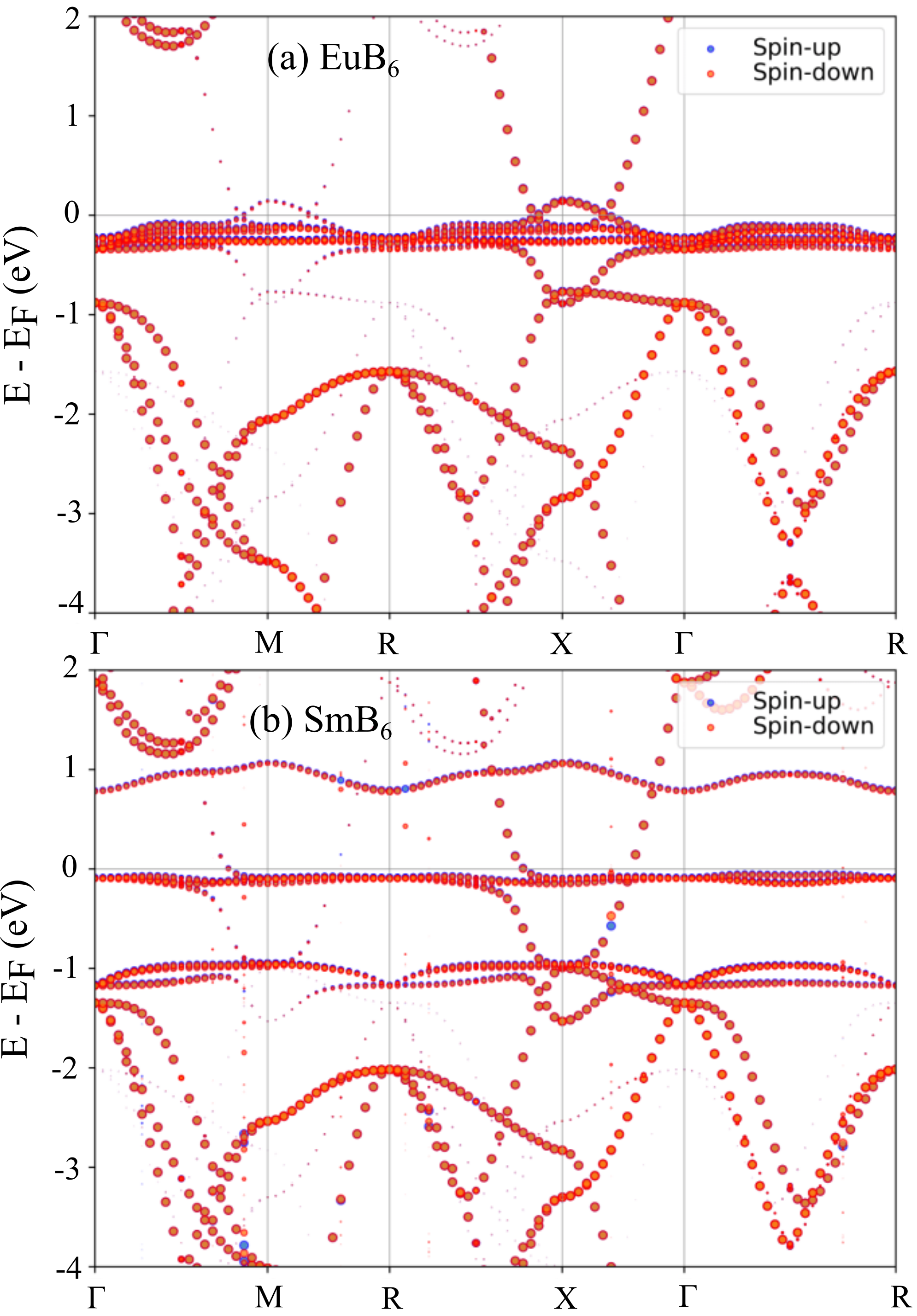}
\captionof{figure}{Unfolded band structures for the AFM configuration. Panel (a) shows the results for EuB$_6$, while (b) illustrates SmB$_6$.}
\label{fig:bands_afm}
\end{center}

One way to probe the presence and nature of the exchange field is by observing its effect on nearby nonmagnetic atoms, which can develop induced magnetic moments in response to the local field. This is precisely what has been experimentally observed in systems like EuTe$_2$, where Te atoms acquire small moments due to the exchange field generated by neighboring Eu atoms~\cite{Sun2023}. In our case, the B atoms serve a similar role: they interact with the exchange field produced by the surrounding Eu atoms and respond accordingly. In the FM configuration, where all Eu spins are aligned, the B atoms experience an equivalent exchange field and develop small magnetic moments, all antiparallel to the aligned Eu spins. In contrast, in the AFM case, the opposing Eu sublattices cancel out the exchange field at each B site, resulting in vanishing local magnetic moments. This indicates that, in both LRO configurations, the exchange field is felt equivalently by every atom, as evidenced by the identical magnetic response of all B atoms.

Comparing with the experimental results discussed in the Introduction~\cite{Donizeth2023,Chen2018}, as all atoms in the LRO phases experience an identical magnetic environment, they should have the same electronic properties. This result is in agreement with the presence of a single XANES/XAS peak, indicating uniform and highly symmetric magnetic surroundings.

\subsection{Polymorphous configuration}

Within our model of the polymorphous paramagnetic phase, the magnetic moments of each Eu or Sm atom are randomly distributed in the lattice (local symmetry is broken), with the constraint that the sum of all moments add up to zero (global symmetry is preserved). Having such a random distribution of atoms leads to the existence of different local environments where the neighboring atoms have different spin states. As the total number of Eu first neighbors is six, the number of neighbors with spin up (or down) can range from zero to six. 

As we have a relatively small supercell, with eigth Eu/Sm atoms, we observe two inequivalent Eu/Sm sites, as shown in Fig. \ref{fig:clusters} (a), each exhibiting distinct electronic properties. This differentiation stems exactly from the variations in the local spin environment, which generate a non-symmetric exchange field $\boldsymbol{B}_{xc}$ and induce symmetry breaking in the electronic structure of the rare-earth atoms. The non-equivalence of $\boldsymbol{B}_{xc}$ can be seen from spatially varying magnetization on the B atoms, in contrast with the LRO configurations, where uniform magnetic environments induces identical magnetic moments on each B atom, as discussed.

The $R_s$ sites (where $R$ stands for rare-earth and $s$ for symmetric) are surrounded by first-neighbor rare-earth atoms with the same spin orientation, forming a well-defined spin cluster. In contrast, the asymmetric $R_a$ sites of spin up (or down) have a mixed first-neighbor spin environment: each $R_a$ atom always has two first neighbors with spin up (or down) and four with spin down (or up), breaking the local spin symmetry. As a result, the exchange field $\boldsymbol{B}_{xc}$ is fully symmetric at the $R_s$ sites, while at the $R_a$ sites it presents a lower symmetry due to this spin mixture in the local environment. Fig.~\ref{fig:clusters}(c) and (d) clarify the different local environments that lead to different local exchange fields. This differentiation gives rise to distinct electronic signatures, indicating that $R_s$ and $R_a$ sites are electronically unequivalent.
\begin{center}
    \includegraphics[scale=0.38]{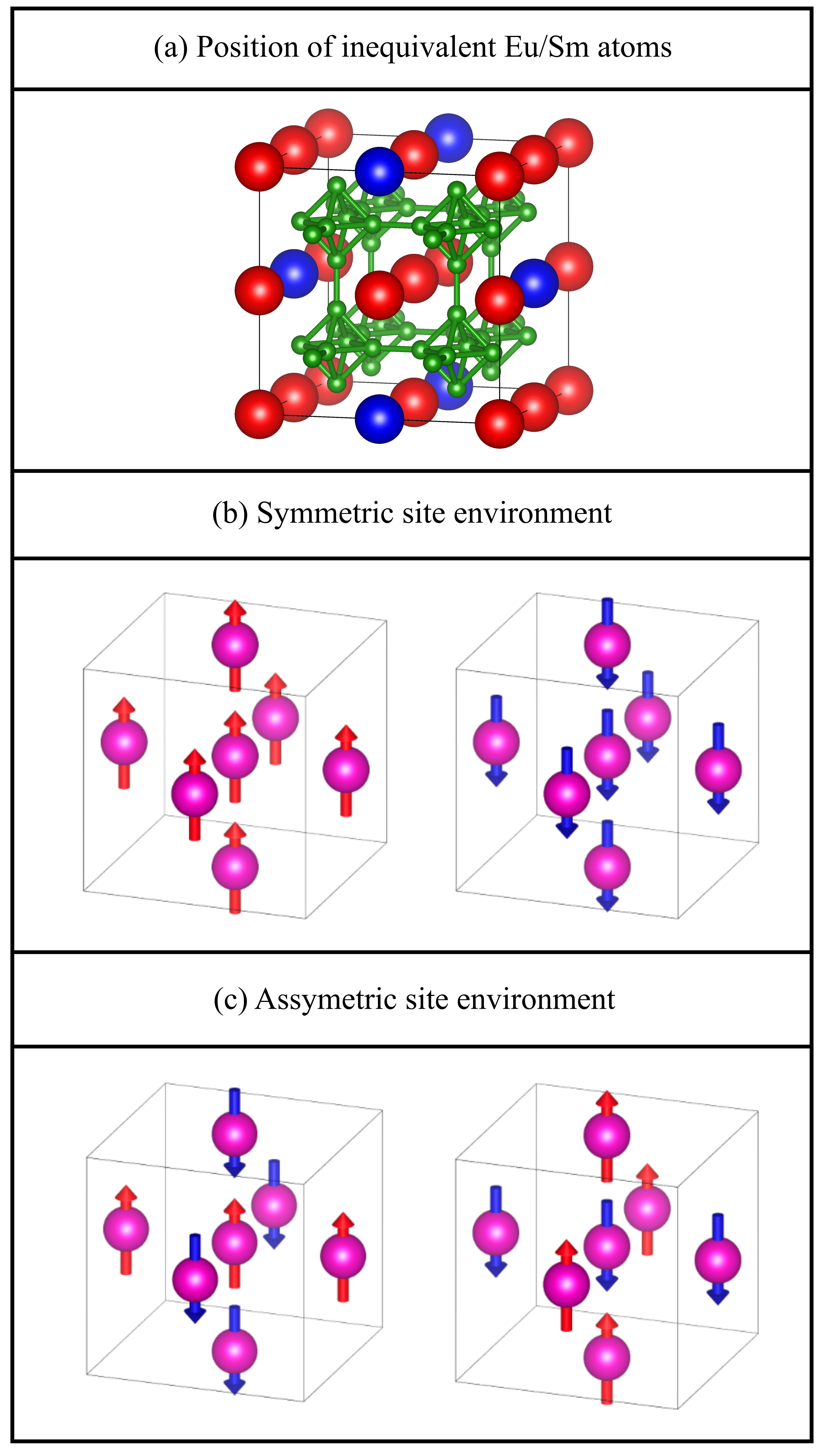}
    \captionof{figure}{(a) Positions of inequivalent Eu/Sm atoms in the PM phase: red atoms denote asymmetric ($a$) sites, while blue atoms correspond to symmetric ($s$) sites. (b) Local environment around a symmetric atom. (c) Local environment around an asymmetric atom.}
    \label{fig:clusters}
\end{center}

The unfolded band structure for the PM configuration is presented in Fig.~\ref{fig:bands_pm}. In the case of SmB$_6$, a broadening on the $f$ levels is visible. This broadening originates from spin polymorphism and is directly linked to the emergence of two distinct atomic environments within the crystal. This broadening becomes evident when comparing the $f$ bands in the FM and AFM band structures with those in the PM configuration.

\begin{center}
\includegraphics[scale = 0.515]{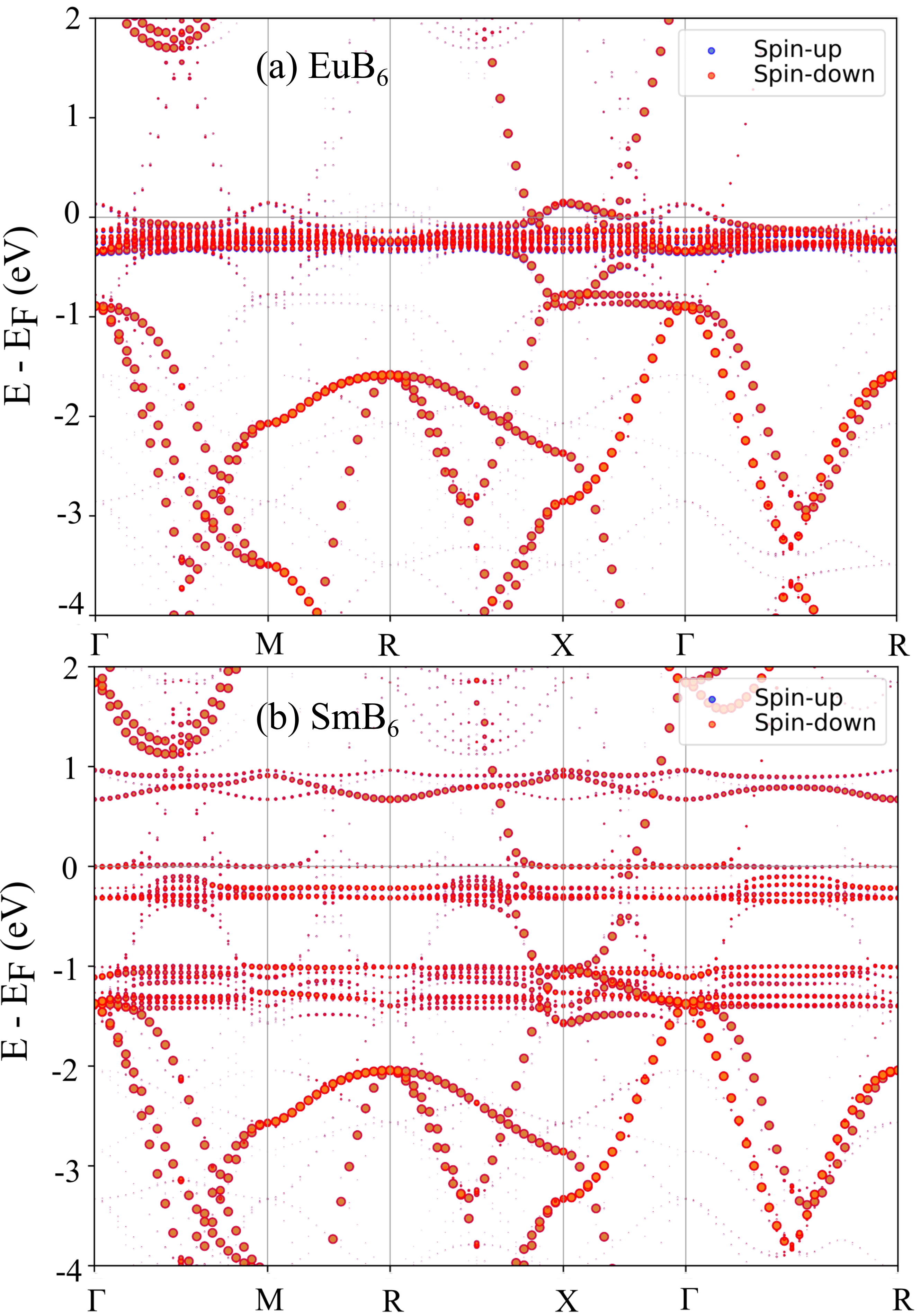}
\captionof{figure}{Unfolded band structures for the PM configuration. Panel (a) shows the results for EuB$_6$, while (b) illustrates SmB$_6$.}
\label{fig:bands_pm}
\end{center}

An analysis of the Projected Density Of States (PDOS) confirms that this broadening arises from the presence of inequivalent rare-earth sites, providing a clear signature of spin-driven symmetry breaking, which in turn leads to electronic symmetry breaking. This effect is explicitly illustrated in Fig.~\ref{fig:dos_eu1_eu2}. The PDOS is normalized per atom and projected onto the symmetric and asymmetric sites within the supercell. Within each site type, the DOS is identical for all atoms belonging to that group. The states presented are predominantly of $f$ character from the rare-earth atoms, and it is evident that the symmetric and asymmetric sites display distinct behaviors.

EuB$_6$ also shows a site-dependent character in its $f$ states, as seen in the projected local DOS, but without any observable $f$-level broadening. While the atoms are no longer fully equivalent, the distinction between sites is more subtle. Nevertheless, this inequivalence is supported by clear differences in the charge density distribution and in specific features of the electronic spectrum, as we shall see for both materials.

\begin{center}
    \includegraphics[scale=0.275]{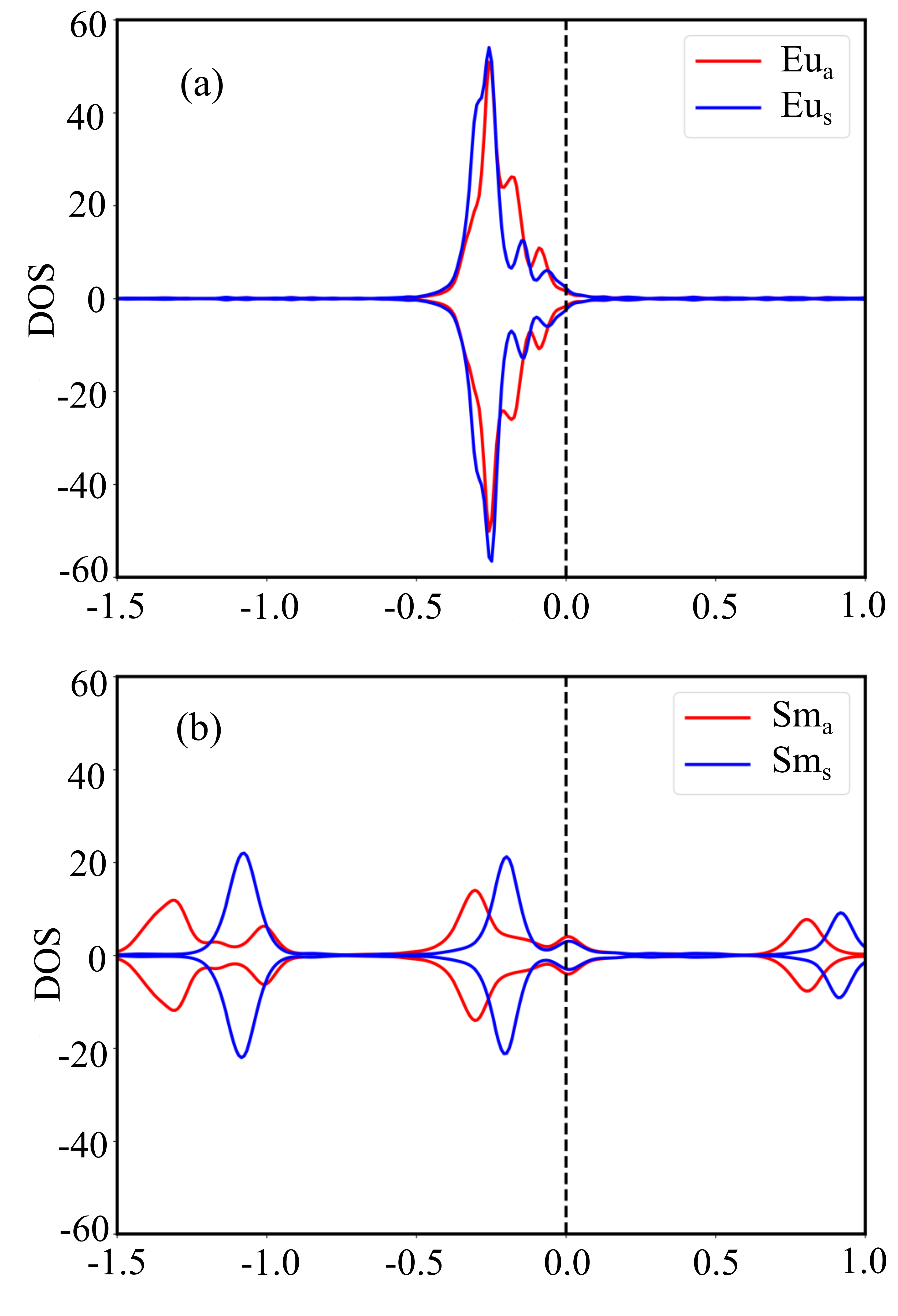}
    \captionof{figure}{Projected density of states (DOS). Both panels show the DOS projected onto symmetric (blue) and asymmetric (red) atomic sites, normalized per atom. Panel (a) presents the projected DOS for EuB$_6$, while panel (b) shows the projected DOS for SmB$_6$.}
    \label{fig:dos_eu1_eu2}
\end{center}

We conducted an analysis (fig. \ref{fig:spectrum_symmetry_gamma}) of the KS eigenvalue spectrum at the $\Gamma$ point, as this point preserves all the symmetry operations of the crystal, providing a clear view of the symmetry-influenced electronic structure. We observe multiple triply degenerate states, for both EuB$_6$ and SmB$_6$, derived from the crystal field splitting of the $f$ orbitals. We evaluated the orbital contribution for $R_s$ and $R_a$ sites into these triply degenerate $f$-levels.

For the symmetric atoms, the projections onto the threefold-degenerate states are uniform across the entire degenerate subspace. This indicates that each symmetric site contributes equally to all the states within the subspace, consistent with the underlying crystal symmetry. In contrast, for the asymmetric atoms, the projections are unevenly distributed across the same subspace, meaning that different states within the degenerate manifold receive different contributions from a given asymmetric site. This non-uniform distribution does not stem from structural inequivalence, but rather from the magnetic symmetry breaking present in the system. As a result, while the global degeneracy of the states is preserved, their spatial distribution becomes sensitive to the magnetic environment, with the asymmetric atoms introducing a site-dependent modulation in the character of the electronic wavefunctions within the degenerate subspace. This effect is depicted in figure \ref{fig:spectrum_symmetry_gamma}. Notably, the effect of the symmetry-broken exchange field on our spectrum appears to break a different type of symmetry — one related to the spectral weight of each state.

\begin{center}
    \includegraphics[scale=0.46]{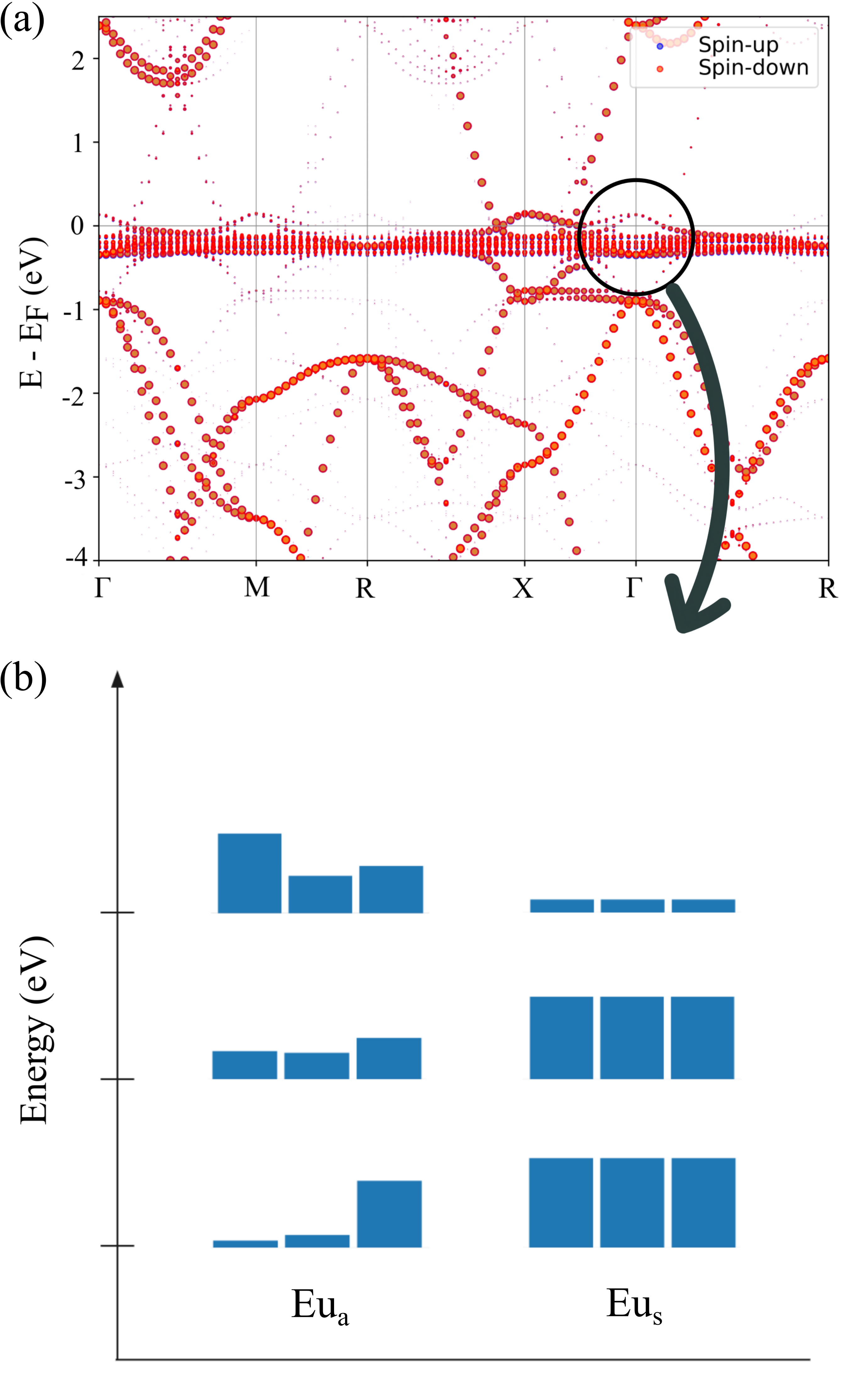}
    \captionof{figure}{(a) Band structure for the paramagnetic (PM) phase of EuB$_6$. The circle highlights Eu $f$ orbitals at the $\Gamma$ point near the Fermi energy. (b) Schematic diagram of three triply degenerate bands at $\Gamma$ with strong Eu $f$ character. For the symmetric Eu atom the $f$ contribution is equal across the three orbitals; in the asymmetric case the contributions are different, revealing an electronic symmetry breaking on top of a symmetric crystal field.}
    \label{fig:spectrum_symmetry_gamma}
\end{center}

The total charge density on the rare earth elements is also affected by the different exchange field environments. By plotting the charge density difference between the polymorphous PM configuration and a LRO configuration, such as the FM phase ($\Delta\rho(\boldsymbol{r}) = \rho_{_{PM}}(\boldsymbol{r}) - \rho_{_{FM}}(\boldsymbol{r})$), one can observe how polymorphism affects the system’s charge distribution. The charge density, $\rho$, behaves differently for each atom, as shown in figure \ref{fig:charge_difference}. Specifically, symmetric atoms appear to lose charge, while asymmetric atoms exhibits both charge gain and loss for both EuB$_6$ and SmB$_6$. Similar charge density analyses have been recently correlated to superconducting materials \cite{Liu2025}.
\begin{center}
    \includegraphics[scale=0.232]{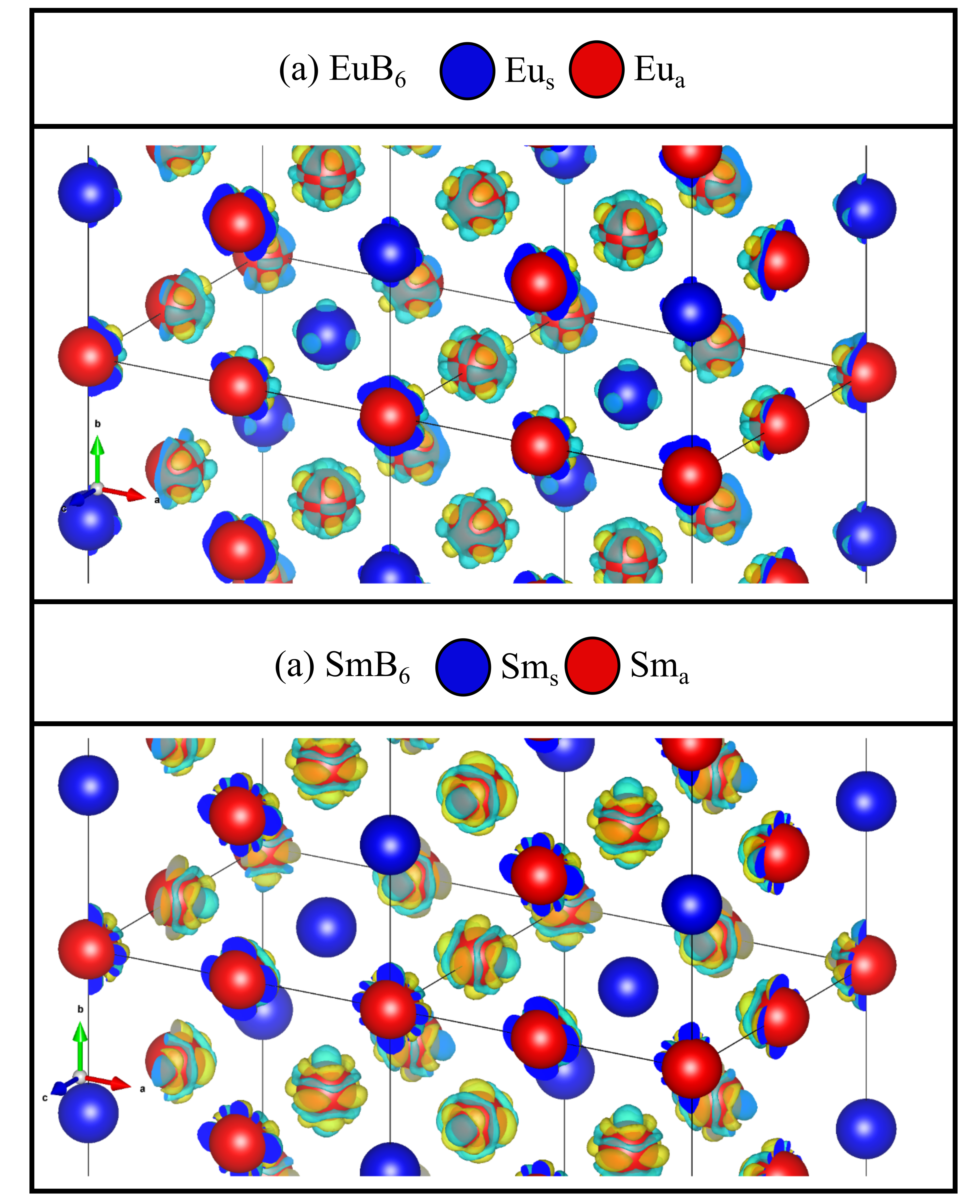}
    \captionof{figure}{\justifying Charge density difference among the PM and FM phases for (a) EuB6 and (b) SmB6. Cyan regions indicate loss of charge density, while yellow regions indicate gain of charge density. The isosurface values are $0.0002\: e/$\AA$^3$ for EuB$_6$ and $0.001\: e/$\AA$^3$ for SmB$_6$. The charge density difference pattern is distinct for the symmetric and asymmetric atoms.}
    \label{fig:charge_difference}
\end{center}

The analysis provides clear evidence that the exchange field acting on a given rare-earth atom can significantly modify its electronic properties. It is important to emphasize that no structural distortion occurs in these systems: all spins occupy perfectly symmetric sites. This is evidenced by the persistence of triply degenerate states in the electronic structure at the $\Gamma$ point. The example presented here adds to the previously reported cases in which symmetry breaking has been identified in quantum materials.

To further test the stability of our conclusions, we performed DFT+$U$ calculations in which different on-site interactions were assigned to the inequivalent rare-earth sites, thereby artificially enhancing their electronic asymmetry in an attempt to nudge the system toward a mixed-valence configuration. While this procedure initially amplifies the differences between symmetric and asymmetric sites, using the resulting charge density as the starting point for an $r^2SCAN$ calculation without on-site corrections drives the system back to the subtle distinctions reported in our work. No evidence of mixed valence was found, reinforcing the robustness of the picture described here.

Calculations were also performed including Spin-Orbit Coupling (SOC) to verify the robustness of our results, since SOC is important for both elements, Eu and Sm. The main conclusions of our paper remained unchanged: properties symmetry breaking arises from the presence of distinct magnetic environments—i.e., rare earth atoms feel different exchange fields in the polymorphic configuration. In our approach, we applied SOC on the PM phase, which naturally yields different exchange fields for different rare-earth sites. The results of the SOC calculations are explicitly shown in the supplementary materials.

\section{Experimental validation}

The results shown here may also provide an alternative explanation for the recent experiments reported in Refs.~\cite{Donizeth2023,PhysRevB.105.195134,Chen2018}, as discussed in the introduction. XAS, and consequently XANES, are multiple scattering techniques \cite{RAVEL2005118} that measure the absorption cross-section as a function of photon energy. When the photon energy matches the binding energy of a deep core electron, the probability of absorption and subsequent promotion of this electron to an unoccupied state increases sharply. This probability is sensitive to the local electronic and atomic structure around the absorbing atom, meaning that variations in the spectral peaks reveal differences in its local environment \cite{RAVEL2005118}.

In the present case, spin symmetry breaking in the PM phase generates distinct magnetic environments for the rare-earth atoms. Even though the crystal structure remains symmetric, the local spin degrees of freedom differ between these sites, producing non-equivalent magnetic surroundings. These magnetic differences directly lead to distinct electronic environments, since the exchange-correlation magnetic field $\boldsymbol{B}_{xc}$ - which enters explicitly in the Kohn–Sham Hamiltonian - is different for each site.

Because $\boldsymbol{B}_{xc}$ modifies the Hamiltonian, the eigenvalue problem that defines the electronic structure is also different for each environment. As the XANES spectrum is directly related to the electronic structure of the material — which are exactly the eigenvalues — the site-dependent $\boldsymbol{B}_{xc}$ should result in measurable differences in the spectra.

Therefore, the double-peak structure observed in XAS/XANES can be attributed to the coexistence of symmetric and asymmetric atomic sites, each characterized by a distinct $\boldsymbol{B}_{xc}$ and, consequently, a distinct electronic structure. In EuB$_6$, this scenario is realized experimentally under pressure, while in SmB$_6$ it occurs already at ambient conditions. In both cases, the non-equivalent magnetic and electronic environments are captured by the multiple scattering processes in XAS/XANES, producing the observed spectral splitting.

\section{conclusions}

Spin symmetry breaking (SB) emerges as a central ingredient for accurately describing both quantum hexaborides, EuB$_6$ and SmB$_6$. Our total energy calculations show that the monomorphous nonmagnetic (NM) configurations are highly unstable, lying significantly above all spin SB configurations. In both NM and LRO magnetic phases, all rare-earth atoms are symmetry-equivalent, as they experience the same exchange field. This global magnetic order suppresses any spatial differentiation of local electronic environments. In contrast, in the PM phase modeled via special quasirandom structures (SQS), the exchange field $\boldsymbol{B}_{xc}$ becomes spatially inequivalent. As a result, the rare-earth atoms no longer experience identical local magnetic environments, which leads to the breaking of both electronic and magnetic symmetries. Our calculations reveal clear signatures of this local symmetry breaking: the projected density of states (PDOS) shows the emergence of inequivalent atomic contributions; the charge density difference between FM and PM configurations reveals spatially varying charge behavior; and the spin-resolved spectra display distinct contributions. All these indicators confirm that symmetry breaking in the PM phase manifests in both spin and charge channels. Notably, this effect is more pronounced in SmB$_6$. Finally, our theoretical predictions align well with recent XAS and XANES experiments on EuB$_6$ and SmB$_6$, which report the coexistence of multiple atomic environments. These observations can now be interpreted as a natural consequence of spin symmetry breaking. A compelling way to test this interpretation would be to apply an external magnetic field within the paramagnetic regime. If the double-peak structure observed in the XAS/XANES spectra vanishes under such a field, it would indicate that magnetic alignment restores uniformity in the local environments, further validating the link between symmetry breaking, magnetic disorder, and correlation in these materials.

\section{Aknowledgements}

The work in Brazil was supported by Brazilian agencies FAPESP (processes 2023/03493-0 and 2023/09820-2) and CNPq. The work in the USA was supported by the National Science Foundation under Grant No. CHE-2533416. We thank CENAPAD-SP and LNCC (Santos Dumont Supercomputer) for computer time. The authors thank Ricardo Donizeth dos Reis and Leonardo Kutelak for fruitful discussions on the subject, and Márcia Fantini for valuable conversations regarding multiple scattering experiments.

\hfill
\pagebreak

\FloatBarrier
\pagebreak

\bibliographystyle{apsrev4-1} 

\bibliography{ref.bib}

\end{document}